\documentclass[twocolumn,showpacs,amssymb,aps]{revtex4}
\usepackage{amsmath}
\usepackage{amssymb}
\usepackage{graphicx}
\usepackage{url}

\def\be{\begin{equation}}
\def\ee{\end{equation}}
\def\bea{\begin{eqnarray}}
\def\eea{\end{eqnarray}}
\def\ba{\begin{array}}
\def\ea{\end{array}}
\def\ben{\begin{enumerate}}
\def\een{\end{enumerate}}

\allowdisplaybreaks[3]

\begin{document}

\title{Search for Cosmic Axions using an Optical Cavity}

\author{A.C.Melissinos \footnote{e-mail: meliss@pas.rochester.edu}}

\affiliation{ Department of Physics and Astronomy, University of
Rochester, Rochester, NY 14627, USA}
\date{\today}


\begin{abstract}

A high finesse optical cavity can be used to search for cosmic
axions in the mass range $10^{-6}< m_a < 10^{-4}$ eV. Either a
two-arm or a single-arm cavity is suitable, and in either case the
signal appears as resonant sidebands imposed on the carrier.
Assuming for the local axion density the usual figure of $\rho_a =
500$ Mev/cm$^3$ \cite{Turner} , the KSVZ axion line \cite{Kim} $g_{a
\gamma \gamma}/m_a = 0.4\ \rm{GeV}^{-2}$, can be reached over the
full mass range in a one year search.

\end{abstract}


\maketitle

 The existence of axions, light pseudoscalars particles,
was postulated thirty years ago \cite{PQ, Weinberg, Wilczek} to
explain why the strong interactions conserve $P$ and $CP$ in spite
of the fact that the QCD Lagrangian does not \cite{Kim,DFS}. Axions
remain an attractive candidate for the cold dark matter of the
universe \cite{Bradley, Sikivie}. A detailed review of axion theory
can be found in \cite{Bradley}. As a result of their gravitational
attraction and very weak interaction with ordinary matter, axions
are expected to condense into galactic halos. The local axion
density is estimated to exceed their average density by a factor of
$\sim 10^5$ \cite{Turner}. We will use units of $\hbar = c = 1$
throughout unless otherwise indicated.

Axions couple to two photons through the triangle anomaly and the
effective action density can be written \cite{Sikivie, Sikivie2007}

\begin{equation}
{\cal L} = \frac{1}{2} \left[ E^2 -B^2 \right] + \frac{1}{2}
 \left( \frac{\partial \phi}{\partial t} \right)^2 - \frac{1}{2}
 \left( \vec{\nabla} \phi \right)^2 -\frac{1}{2} m^2_a \phi^2 -g \vec{E} \cdot \vec{B} \phi
\end{equation}

\noindent $\vec E$ and $\vec B$ are the electric and magnetic field,
and $\phi, m_a$ the axion field and axion mass. The coupling of the
axion to two photons is designated by $g$ and is proportional to the
axion mass. In order of magnitude

\begin{equation}
g \equiv g_{a \gamma\gamma} = \frac{1}{\Lambda} \simeq
\frac{\alpha}{\pi} \frac{m_a}{m_\pi f_\pi}
\end{equation}

\noindent with $m_{\pi}, f_{\pi}$ the pion mass and decay constant,
$m_{\pi}f_{\pi} \sim 10^{-2}\ \rm{GeV}^{2}$. In all axion models the
product of the inverse coupling constant, $\Lambda$(GeV) and axion
mass is constant with $\Lambda m_a \sim 1\ \rm{GeV}^2$. The
classical equations of motion for the fields derived from Eq.(1) are

\begin{eqnarray}
&&\vec{\nabla} \cdot \vec{E} = g \vec{B} \cdot \vec{\nabla} \phi\\
\noalign{\vskip 4pt}%
&&\vec{\nabla} \times \vec{B} - \frac{\partial \vec{E}}{\partial t}
= g \left[ \vec{E} \times \vec{\nabla} \phi - \vec{B}
\frac{\partial \phi}{\partial t} \right]\\
\noalign{\vskip 4pt}%
&&\left[ \frac{\partial^2}{\partial t^2} - \nabla^2 \right] \phi +
{m_a}^2 \phi =- g \vec{E} \cdot \vec{B}
\end{eqnarray}

Over the past two decades several experiments have searched for
cosmic axions \cite{Semertzidis, Wuensch, Hagman, Bradley}, and for
axions produced in the sun \cite{Lazarus, CAST}. There have also
been efforts to observe axion production using laser beams
\cite{LLNL, Cameron, Ruoso, PVLAS}. Dark matter candidate axions are
expected in the mass range $10^{-3} < m_a < 10^{-6}$ eV
\cite{KolbTurner, Bradley}, with correspondingly weak couplings to
the em field. The most sensitive searches for micro-eV axions in the
galactic halo are based on the conversion of axions to microwave
photons in a static magnetic field. The converted photons are
detected in a cavity which is resonant at the frequency
corresponding to the axion mass \cite{Bradley}.

Here we propose an analogous process where the axions are absorbed
(but also emitted) by (from) an optical field of frequency
$\omega_0$, typically in the visible. Therefore sidebands
$\omega_{\pm} = \omega_0 \pm \omega_a$ appear on the carrier,
displaced by the axion frequency $\omega_a = E_a \simeq m_a$. For
this process to be efficient, the sidebands must resonate in the
optical cavity. We discuss later how this is achieved in practice.

We start from Eqs.(3-5) and designate the carrier fields by $\vec
E_0$, $\vec B_0$ and the sideband fields by $\vec E_{\pm}$, $\vec
B_{\pm}$ for the upper and lower sideband respectively;
\begin{equation}
\vec{E} = \vec{E}_0 + \vec{E}_{+} +\vec{E}_{-} \qquad {\rm{and}}
\qquad \vec{B} = \vec{B}_0 + \vec{B}_{+} +\vec{B}_{-}
\end{equation}

\noindent The fields $\vec{E}_{\pm},\vec{B}_{\pm}$ are of order
$g\phi_a \sim 10^{-21}$ as compared to the fields
$\vec{E}_0,\vec{B}_0$. The carrier is a standing wave in a cavity of
length $L$ along the x-axis

\begin{eqnarray}
\vec{E}_0 &=& A_0(t) \sin (k_0 x) e^{-i \omega_0 t}\hat{u_z}
\nonumber \\
\vec{B}_0 &=& A_0(t) \cos (k_0 x) e^{-i( \omega_0 t
-\pi/2)}\hat{u_y}
\end{eqnarray}

\noindent where $\omega_0 = k_0 = n_0(\pi/L)$. We seek solutions
where the sideband fields are orthogonal to the carrier and are
standing waves in an overlapping cavity of length $L_s$,

\begin{eqnarray}
\vec{E}_{\pm} &=& \pm A_{\pm}(t) \sin (k_{\pm} x) e^{-i \omega_{\pm}
t}\hat{u}_y  \nonumber \\
 \vec{B}_{\pm} &=& \pm A_{\pm}(t) \cos
(k_{\pm} x) e^{-i (\omega_{\pm} t - \pi/2)}\hat{u}_z
\end{eqnarray}

\noindent where $\omega_{\pm} = k_{\pm} = \omega_0 \pm \omega_a$.
The length $L_s$ is adjusted to make one of the sidebands resonant,
i.e. $\omega_{+} = n_{+}(\pi/L_s)$. When the axion frequency
coincides with multiples of the free spectral range of the
overlapping cavity, $\nu_a = \omega_a /2\pi = q/2L_s$ both the upper
and lower sidebands are simultaneously resonant. In Eqs.(8) we have
explicitly indicated the slow variation in time of the amplitudes
$A_{\pm}(t)$. We have also written $A_0(t)$ in Eqs.(7) even though,
in practice, the carrier amplitude remains constant.

The axion field is assumed spatially homogeneous over the dimensions
of the detector

\begin{equation}
\phi(x,t) = \phi_a (e^{-i \omega_a t} + e^{i \omega_a t})/2
\end{equation}

\noindent This assumption is justified because the DeBroglie
wavelength of the axions $\lambda_{DB} = 2\pi/(\beta_a m_a)$ is much
larger than the dimensions of the detector for $m_a < 10^{-3}$ eV.
$\beta_a$ is the velocity of the axions which is that of the virial
velocity of the galaxy $\beta_a \simeq 10^{-3}$. The first term in
Eq.(9) contributes the upper sideband and the second term the lower
sideband.

Introducing Eq.(6) in Eq.(4) and keeping only terms of order
$g\phi_a$ we obtain the wave equation for the sideband fields. We
made use of Eq.(3) and also of $\omega_a<<\omega_0$ to neglect terms
in $\omega_a/\omega_0$. We find

\begin{equation}
\left( \nabla^2 - \partial^2 /\partial t^2 \right) \vec{E}_{\pm} =
\pm g \omega_0 \omega_a \vec{B}_0 \phi_a/2
\end{equation}

\noindent As expected the sideband fields $\vec{E}_{\pm}$ are
directed perpendicular to $\vec {E_0}$ and have time dependence
$e^{-i\omega_{\pm}t}$.  The remaining terms give the wave equation
for the evolution of the carrier field which, to order $g\phi_a$ is
free,  $ \left( \nabla^2 - \partial^2 / \partial t^2 \right)
\vec{E}_0 = 0 $.

We can solve Eq.(10) imposing the boundary conditions for a standing
wave (vanishing electric field at the cavity boundaries). The same
result is obtained by using Eq.(4) directly

\begin{equation}
\vec{\nabla} \times \vec{B}_{\pm} - \frac{\partial
\vec{E}_{\pm}}{\partial t} =
 \pm (i/2)g \omega_a \vec{B}_0 \phi_a e^{\mp i \omega_a t}
 \end{equation}

\noindent Using Eqs.(7,8) the rapid time dependence cancels leading
to

\begin{equation}
\frac{dA_{\pm}}{dt} \sin(k_{\pm}x) = \pm g \omega_{a} A_0 \cos(k_0
x) \phi_a /2
\end{equation}

\noindent We expand $\cos(k_0 x)$ in the modes of the overlapping
cavity, $\cos(k_0 x) = \sum C_l \sin(k_lx)$ with $k_l = l\pi /L_s$.
Hence,

\begin{equation}
\frac{dA_{\pm}}{dt} = C_{\pm}g \omega_{a} A_0 \phi_{a}/2
\end{equation}

\noindent $l_{\pm} = (\omega_{0}\pm\omega_{a})(L_s /\pi)$ and $
C_{\pm} = [1 - \cos(k_a L_s)]/\omega_{a} L_s $.

 The growth of the Amplitude $A_{\pm}(t)$ is
restricted by the losses in the cavity, expressed by the ``quality
factor" $Q$

\begin{equation}
\frac{dA_{\pm}}{dt} = - \frac{\omega_{\pm}}{2 Q} A_{\pm}
\end{equation}

\noindent It follows that in the steady state

\begin{equation}
A_{\pm} = \pm  g \phi _a Q C_{\pm}\frac{\omega_{a}}{\omega_0}A_0
\end{equation}

The configuration of the coupled cavities is shown in Fig.1. The
carrier resonates in $L$, between M1 and M2. The sidebands have
orthogonal polarization to the carrier and are directed by the
(polarizing) beam splitter to mirror 3. The spacing, $L_{s}$,
between M1 and M3, is tuned to the sideband frequency.

\begin{figure}
\centering
\includegraphics[width=0.8\columnwidth]{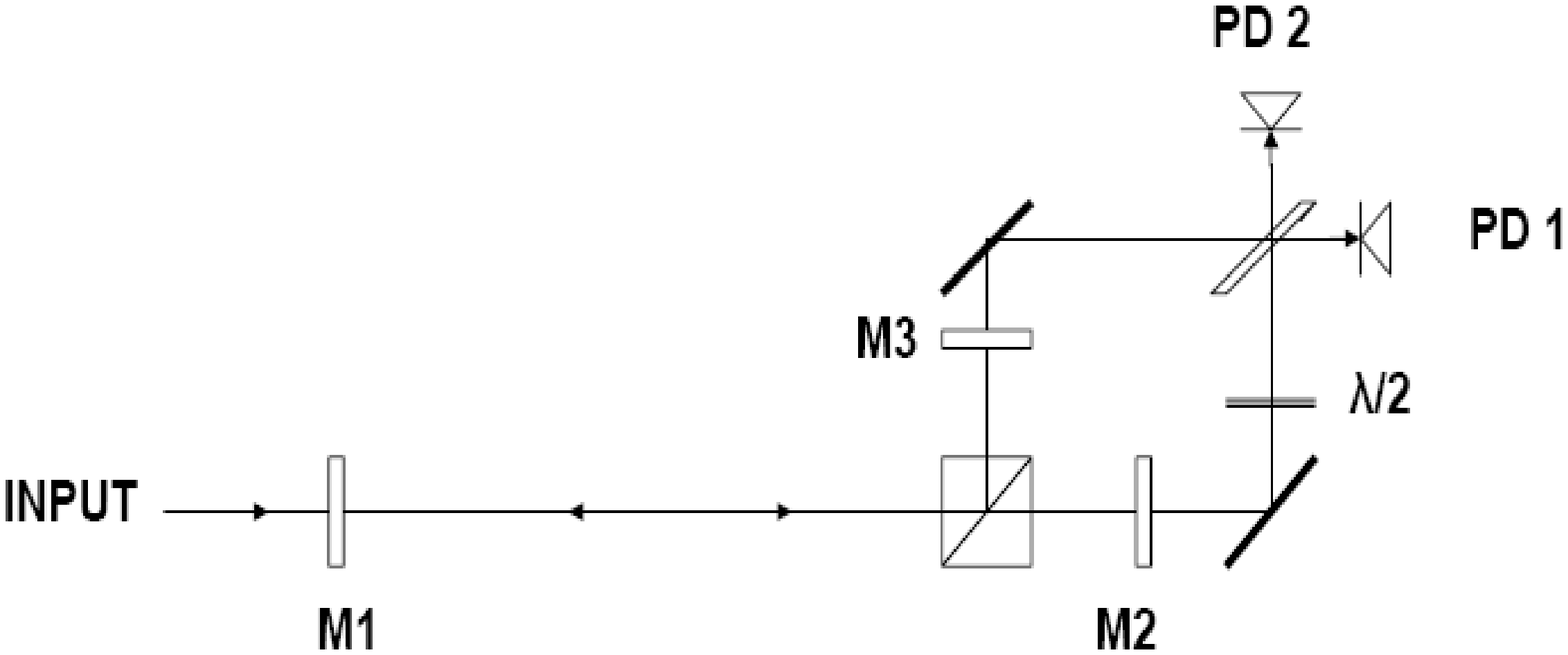}
\caption{Proposed layout of the coupled cavities}
\end{figure}

We define the cavity ``finesse" in the usual way
\begin{equation}
{\cal F} = \pi {\sqrt{r_1 r_2}\over {1 - r_1r_2}}
\end{equation}

\noindent where $r_1,r_2$ are the amplitude reflectivities of the
input and output mirrors and other losses are assumed absent. The
quality factor of the cavity is

\begin{equation}
Q = {\cal F}(2L/\lambda_0)
\end{equation}

\noindent with $L$ the length of the cavity and $\lambda_0$ the
wavelength of the carrier. The free spectral range (fsr) of the
cavity is $\nu_{fsr} = 1/(2L)$ and the FWHM of the cavity resonance
is $\Delta \nu_c = \nu_{fsr}/{\cal F} = \nu_0/Q$. We use $t_1,t_2$
for the amplitude transmissivities of the cavity mirrors, which
satisfy $r^2 + t^2 + A = 1 $, with $A$ the absorption coefficient
(for simplicity we set $A$ = 0). The carrier field circulating in
the cavity is $E^{circ} = E^{in} t_1 /(1-r_1 r_2)$ and the
transmitted fields $E^{out} = t_2 E^{circ}$; $E^{in}$ is the
incident field.

With the above definition of the $Q$ of the optical cavities Eq.(15)
can be written in compact form

\begin{equation}
\frac{A_{\pm}}{A_0} = \pm g\phi_a ({\cal F}/\pi) [1-\cos(k_a L_s)]
\equiv R_{\pm}
\end{equation}

\noindent It is interesting, but not surprising, that the
experimental signal depends only on the product of the axion
coupling and field, which is constant for fixed axion density. This
is also true for the microwave cavity axion searches.

 The axion amplitude $\phi_a$
is related to the axion density $\rho_a$ through

\begin{equation}
\langle \phi^2 \rangle = \rho_a/m_a^2 \qquad\qquad \rm{or}
\qquad\qquad \phi_a = \sqrt{2 \rho_a}/\it{m_a}
\end{equation}

\noindent Using $\rho_a = 0.5\ \rm{GeV/cm^3}$ \cite{Turner} and $
g/m_a = 0.4\ \rm{GeV}^{-2}$ appropriate for the KSVZ model
\cite{Kim} we obtain for the dimensionless quantity $g_a \phi_a =
10^{-21}$. We can expect a finesse ${\cal F} \sim 10^6$
\cite{PVLASB}, so that when the spatial form factor is optimized, we
must be able to detect sidebands of relative amplitude $A_{\pm}/A_0
\sim 10^{-15}$, in order to reach the axion limit.

It is not possible to measure directly the power in the sidebands
for such a weak signal but one must instead measure the amplitude of
the sideband field. This is done by exploiting the coherence between
the carrier and the sidebands and measuring the interference term;
what is referred to as ``homodyne" detection.

The carrier reaching the detector is suppressed by the transmission
$\sigma$ through the polarizer. Thus the photodiode current is

\begin{equation}
I_P = \eta |\sigma A_0 + A_{\pm}|^2
\end{equation}

\noindent and we can set the quantum efficiency $\eta \sim 1$. The
irreducible background is determined by the shot noise fluctuations
in the carrier

\begin{equation}
I_N = \sigma^2 \delta |A_0|^2 = \sigma^2 |A_0|^2 [(4 \pi ^2 \hbar
c/\lambda_0)/P_{in} {\cal F}]^{1/2} /\sqrt {\rm Hz}
\end{equation}

\noindent with $P_{in} \propto |A_0|^2 $ the input laser power. The
signal current is $I_S = 2 R_{\pm} \sigma |A_0|^2 $, and the S/N
ratio

\begin{equation}
\frac{S}{N} = 2 R_{\pm} \frac{1}{\sigma}[(4 \pi ^2 \hbar
c/\lambda_0)/P_{in} {\cal F}]^{-1/2}
\end{equation}

\noindent For $P_{in} = 1\ {\rm W},\ \lambda_0 \sim 10^{-6}\ {\rm
m}, \ {\cal F} = 10^6,\ R_{\pm} = 10^{-15}$ and $\sigma = 4\times
10^{-4}$, the shot noise limited S/N is of order 5 for 1 Hz signal
bandwidth. The signal power is significantly above amplifier noise.
For an amplifier with a $10^{\circ}$ K temperature at 24 GHz, the
noise power is $P_N = 1.4\times 10^{-22}$ W/Hz, whereas the
(homodyned) signal power, for $\sigma = 4 \times 10^{-4}$ is $\sim
2.5\times 10^{-18}$ W. The extinction of the carrier is controlled
by $\sigma$ which is eventually chosen so as to optimize the overall
S/N.

A critical issue is the detection of the high frequency modulation
imposed on the optical carrier. Fast photodiodes with bandwidth of
30 GHz are available. Thus an interesting range of axion masses for
this search would be $10^{-6} < m_a < 10^{-4}$ eV, namely from 240
MHz to 24 GHz. In this frequency range microwave techniques will
have to be used to amplify the weak signal in the photocurrent.

The optical cavity needs to be scanned only over one free spectral
range in the region of the axion frequency $\nu_a$, namely at the
$N^{th}$ free spectral range where $N = \nu_a/\nu_{fsr} = 2 L
\nu_a$. To cover the desired range from $N\ {\rm to}\ N+1$ the
cavity length needs to be changed by
$$ \delta L/L = 1/N  \qquad {\rm or}\qquad \delta L \sim 1/2 \nu_a $$
Assuming a 1 m long cavity, and the range of frequencies of
interest, $\delta L = \pm 25$ cm covers the low mass range, $m_a =
10^{-6}$ eV, whereas for $m_a =10^{-4}\ {\rm eV}, \delta L = \pm
2.5$ mm. Scanning over the entire 25 cm range will result in
repeated resonances if the axion mass is in the upper part of the
search.

The width of the cavity resonance is fixed by the finesse and the
fsr frequency, $\Delta \nu = \nu_{fsr}/{\cal F} =150$ Hz for $L$ = 1
m. The width of the axion line is determined by the random motion of
the axions $\Delta \nu_a = (1/2)m_a \beta ^2 =\linebreak 5 \times
10^{-7} m_a$. This ranges from 120 Hz to 12 kHz. One would scan in
steps of the cavity width at the $N^{th}$ fsr. Namely a total of
${\cal F}$ steps with progressively smaller increments, depending on
the axion mass that is searched for, i.e. $\delta x_{step} = \delta
L(\nu_a)/{\cal F}$. Devoting an integration time of 30 s to each
step would require about a year for the entire search, at a shot
noise limited S/N = 2.5. In this search the carrier frequency is
locked onto the primary cavity and the length of the sideband cavity
is continuously scanned.

The insertion of the polarizing beam splitter in the carrier cavity
will introduce losses and prevent the finesse from reaching the
design figure of ${\cal F} = 10^6$. Instead one can use a single
cavity and consider length settings where both the carrier and the
sidebands are resonant. All previous calculations remain valid, but
now both the upper and lower sidebands are present with equal
amplitudes. The resonance conditions are $ k_0 L/\pi = n_0$ and
$k_{\pm} L/\pi = n_{\pm} $, and therefore
\begin{equation}
\omega_a L/\pi = (n_{\pm} - n_0) = 2p + 1
\end{equation}
\noindent where $p$ is an integer. We impose the odd integer
condition so that the coefficient $C_{\pm} = 2/\omega_a L$, while it
vanishes when $(n_{\pm} - n_0)$ is even.

The drawback of using a single cavity is that the carrier frequency
must be continuously adjusted as the cavity length is scanned. In
practice this is quite feasible except when using a short cavity to
search in the lowest range of axion masses. Since both sidebands
resonate and have opposite (real) amplitudes, mixing with the
carrier gives a null result. In order to detect the signal one must
impose FM sidebands on the carrier and mix the modulated fields with
the signal field.


So far it has been implicitly assumed that the axion field is
coherent  \cite{referee}. While this is true when the axion field
was first created, during the evolution to the present time the
field has retained only partial coherence. The present coherence can
be estimated from the time it takes an axion to traverse its
DeBroglie wavelength \cite{Matsuki}.

\begin{equation}
\tau_{coherence} \sim 2\pi/m_a \beta_a^2
\end{equation}

\noindent Numerically, $\tau_c = 4\times 10^{-3}/[ m_a/ 1\ \mu
{\rm{eV}}]$ s. Such partial coherence times are longer than the
``filling" time required for the optical cavity to reach the signal
level, $ \tau_{acq} = 2Q/\omega_c = 10^{-3}$ s; see Eq.(14).
Therefore the data can be acquired at intervals shorter than the
coherence time of the axion field. Sampling at a rate of $f_s = 2.5$
kHz ($\tau_s = 4\times10^{-4}$ s) will reduce the S/N ratio of each
individual measurement to 0.1. However averaging the $ f_s\times
\Delta T = 7.5\times 10^4$ individual measurements restores the
previously quoted sensitivity, in this case the ratio of the signal
to the fluctuations of the noise.

As long as the acquisition time is shorter than the coherence time
of the axion field, no information is lost. Apart from the
complication of faster sampling, and correspondingly larger data
sets, the statistical accuracy can be recovered by off-line data
processing. Recent studies of axion flows in the vicinity of the
earth \cite{SikivieYang} suggest that the axions form a
Bose-Einstein condensate, and predict coherence times in the order
of seconds, even for axion masses as high as $m_a = 10^{-4}$.

A search for cosmic axions using an optical cavity can reach the
axion limit and cover the entire cosmologically interesting range
$10^{-6}\ {\rm to}\ 10^{-4}$ eV in one year. As compared to the
microwave cavity searches the optical technique has the advantage of
measuring the amplitude (rather than the power) of the induced em
field. Of course, the ``external" magnetic field is significantly
weaker. Another advantage is that with the optical cavity the search
is broad-band and several axion frequencies are queried at each
setting of the cavity length. Finally there are no special
conditions imposed on the geometric dimensions of the cavity and the
tuning is straightforward. Exploiting the (partially) coherent
nature of the axion field using a microwave cavity has been proposed
previously \cite{Matsuki}. In that case the loaded microwave cavity
is tuned to the axion frequency and the axion signal is extracted
from the slow modulation of the cavity power, rather than by the
direct detection of the sideband power as proposed here.

I thank Profs. A. Das, C.R. Hagen, and L. Stodolsky for useful
discussions and comments. In particular I thank Drs. G. Ruoso and M.
Herzberg for critical comments.

\pagebreak

\pagebreak


\begin{thebibliography}{99}

 \bibitem{PQ} R. Peccei and H. Quinn, Phys. Rev. Lett. \bf{38}\rm, 1440 (1977).

 \bibitem{Weinberg} S. Weinberg, Phys. Rev. Lett. \bf{40}\rm, 223
 (1978)

 \bibitem{Wilczek} F. Wilczek, Phys. Rev. Lett. \bf{40}\rm, 279 (1978).

 \bibitem{Kim} J. Kim, Phys. Rev. Lett. \bf{43}\rm, 103 (1979); M.A.
 Shifman, A.I. Vainshtein and V.I. Zakharov, Nucl. Phys. {\rm B166},
 493 (1980).

 \bibitem{DFS} M. Dine, W. Fischler and M. Srednicki, Phys. Letters \bf{104B}\rm,
 199 (1981); A.R. Zhitnitsky, Sov. J. Nucl. Phys. \bf{31}\rm, 260
 (1980).

 \bibitem{Bradley} R. Bradley et al., Rev. Mod. Phys. \bf{75}\rm,
 777 (2003).

 \bibitem{Sikivie} P. Sikivie, Phys. Rev. Lett. \bf{51}\rm, 1415
 (1983); Phys. Rev. \bf{D32}\rm, 2988 (1985).

 \bibitem{Turner} M.S. Turner, Phys. Rev. \bf{D33}\rm, 889 (1986);
 E.J. Gates, G. Gyuk and M.S. Turner, Astrophys. J. Lett.
 \bf{449}\rm, L123 (1995).

 \bibitem{Sikivie2007} P. Sikivie, D.B. Tanner and K. van Bibber, Phys. Rev. Lett. {\bf 98},
 172002 (2007), arXiv:hep-ph/07011198 (2007).

 \bibitem{Semertzidis} S. De Panfillis et al., Phys. Rev. Lett.
 \bf{59}\rm, 839 (1987).

 \bibitem{Wuensch} W.S. Wuensch et al., Phys. Rev. \bf{D40}\rm, 3153
 (1989).

 \bibitem{Hagman} C. Hagmann et al., Phys. Rev. \bf{D42}\rm, 1297
 (1990).

 \bibitem{Lazarus} D.M. Lazarus et al., Phys. Rev. Lett. \bf{69}\rm, 2333 (1992);
  Y. Inoue et al., Phys. Letters {\bf B668}, 93 (2008) arXiv:0806.2230 [astro-ph].

 \bibitem{CAST} K. Zioutas et al. (CAST Collaboration), Phys. Rev.
 Lett. \bf{94}\rm, 121301 (2005); E. Arik et al. (CAST
 Collaboration), J. Cosmol. Astropart. Phys. JCAP02,008 (2009) arXiv:0810.4482 [hep-ex] (2008).

 \bibitem{LLNL} K. van Bibber et al., Phys. Rev. Lett. \bf{59}\rm,
 759 (1987).

 \bibitem{Cameron} R. Cameron et al., Phys. Rev. \bf{D47}\rm, 3707
 (1993).

 \bibitem{Ruoso} G. Ruoso et al., Z. Phys. \bf{C56}\rm, 505 (1992)

 \bibitem{PVLAS} E. Zavattini et al. (PVLAS Collaboration), Phys.
 Rev. Lett. \bf{96}\rm, 110406 (2006); E. Zavattini et al. (PVLAS Collaboration), Phys.
 Rev. \bf{D77}\rm, 032006 (2008); A. S. Chou et al. (GammeV Collaboration), Phys.
 Rev. Lett. \bf{100}\rm, 080402 (2008); C.Robilliard et al, Phys. Rev. Lett. \bf{99}\rm, 190403
 (2007); A.Afanasev et al. (LIPPS Collaboration) Phys. Rev.Lett.
 \bf{101}\rm, 120401 (2008).

 \bibitem{KolbTurner} E. Kolb and M. S. Turner, {\it{The Early
 Universe}}, Addison-Wesley, Reading MA, 1990.

\bibitem{PVLASB} The PVLAS collaboration has achieved a finesse
${\cal F}= 10^5$; private communication from Dr. G.Ruoso.

 \bibitem{referee} I thank one of the referees for raising the issue of the
 coherence of the axion field.

 \bibitem{Matsuki} S. Matsuki, I.Ogawa and K. Yamamoto, Physics
 Letters \bf{B336}\rm, 573 (1994).


 \bibitem{SikivieYang} P.Sikivie and Q.Yang, arXiv:0901.1106 [hep-ph]
 (2009); see also related articles, P. Sikivie and J. R. Ipser, Physics Letters
 \bf{B291}\rm, 288 (1992); L. D. Duffy and P. Sikivie,
 arXiv: 0805.4556v1 [astro-ph].

\end{thebibliography}
\end{document}